# Digital Nudges Using Emotion Regulation to Reduce Online Disinformation Sharing

HARUKA NAKAJIMA SUZUKI [1,a]    MIDORI INABA [1]

*Abstract*: Online disinformation often provokes strong anger, driving social media users to spread it; however, few measures specifically target sharing behaviors driven by this emotion to curb the spread of disinformation. This study aimed to evaluate whether digital nudges that encourage deliberation by drawing attention to emotional information can reduce sharing driven by strong anger associated with online disinformation. We focused on emotion regulation, as a method for fostering deliberation, which is activated when individuals' attention is drawn to their current emotions. Digital nudges were designed to display emotional information about disinformation and emotion regulation messages. Among these, we found that distraction and perspective-taking nudges may encourage deliberation in anger-driven sharing. To assess their effectiveness, existing nudges mimicking platform functions were used for comparison. Participant responses were measured across four dimensions: sharing intentions, type of emotion, intensity of emotion, and authenticity. The results showed that all digital nudges significantly reduced the sharing of disinformation, with distraction nudges being the most effective. These findings suggest that digital nudges addressing emotional responses can serve as an effective intervention against the spread disinformation driven by strong anger.

## 1. Introduction

Many social media users are hesitant to spread content they deem inaccurate [1]; however, false news spreads more widely than true news [2]. The reason online disinformation spreads is that it is difficult to judge its authenticity. Disinformation is defined by the European Commission as "verifiably false or misleading information that is created, presented, and disseminated for economic gain or to intentionally deceive the public, and may cause public harm" [3]. However, disinformation is not only false but also includes fabricated information blended with facts [4]. Some disinformation is intended to provide correct information that is manipulated as a legitimate method to expose racism [5]. The 2016 Brexit and U.S. election disinformation campaigns were designed to sow mistrust, confusion, and sharpen existing sociocultural divisions using nationalistic, ethnic, racial, and religious tensions [6].

Online disinformation aims to exploit social unrest and cognitive biases to generate extreme anger and suspicion [7]. The most widespread content in disinformation campaigns is that which influences people's emotions and encourages feelings of superiority, anger, or fear [6]. Regarding judgments of accuracy, emotions make people perceive fake news as accurate [8]. Furthermore, people primarily pass on information that evokes an emotional response, irrespective of its truth value [9]. This tendency is particularly observed when strong anger is evoked by disinformation [10]. Not only is anger highly contagious [11], but it also promotes a preference for retributive information and solutions [12]. Anger provokes groups to pursue harmful goals, encouraging them to concentrate on less meaningful objectives or goals that are opposed to their flourishing [13]. For many users, negative emotions caused by social comparison paralyze their ability to pursue values and may not lead to their best behavior [13].

Although refutation strategies, educational interventions, boosts, and nudges are typical user intervention measures to prevent the spread of online disinformation [14], their effectiveness may be limited. Disinformation that is difficult to judge, coupled with strong anger, makes these measures insufficient. First, although refutation strategies may reduce anger [15], they are post-exposure measures that do not influence sharing behaviors during exposure. Second, educational interventions and boosts may be ineffective against the spread of disinformation due to anger because strong anger diminishes their impact [10]. Strong anger influences sharing intentions more than authenticity [10], suggesting that the ability to judge authenticity is not activated at the moment of sharing. These differences stem from the cognitive processes underlying emotion and truth judgment. Bago et al. used dual-process theory [16], [17], which distinguishes between intuitive and deliberative cognitive processes, and found that when people rely on intuition instead of deliberation, they cannot discern true news from false [18]. Anger promotes intuitive cognitive processes [19], [20] and inhibits deliberation. Thus, strong anger may impede the ability to judge authenticity, leading to the easier spread of disinformation.

This study aimed to evaluate whether digital nudges that encourage deliberation by drawing attention to emotional information can reduce sharing driven by strong anger associated with online disinformation. Digital nudging uses user-interface design elements to guide behavior in digital environments [21]. We focused on conventional digital nudge methods that intervene directly in users' sharing behaviors. Since a pop-up window is displayed at the moment disinformation is shared, the timing is more effective than refutation strategies. However, conventional digital nudges typically draw attention to the accuracy of

1 Institute of Information Security, Yokohama, Kanagawa 221-0835, Japan
a) dgs224101@iisec.ac.jp

information [1], [22] or introduce a simple pause [23], [24]. These methods may be insufficient to counter the strong anger that drives the spread of disinformation. We propose a novel approach that draws user's attention to the strong anger they recognize in disinformation. Since reliance on emotion over reason increases judgements errors [8], deliberation that enhances responsiveness to reason [25] may lead to more reflective judgments [26]. To encourage deliberation, we focused on emotion regulation [27], which is triggered when people's attention is drawn to their emotions [28]. Strong negative emotions derail deliberative processes, so it is critical to help moderate internal states and users self-regulate emotions [29]. We developed digital nudges incorporating emotion regulation to intervene when users share disinformation and compared their effectiveness to existing platform-mimicking nudges. Our findings show that distraction nudges, which combined disinformation's emotional information with messages encouraging users to imagine followers smiling when they view the posts, were most effective in reducing the sharing of disinformation. Raising awareness of others and the social impact of sharing disinformation may effectively decrease its spread. To the best of our knowledge, this study is the first to focus on strong anger at the moment of sharing disinformation as a strategy to reduce its spread. The findings provide a new criterion for preventing disinformation propagation by using emotional components rather than accuracy. This approach is useful for disinformation that blends truth with falsehood and may also prevent the formation of false beliefs tied with emotions.

The remainder of this paper is organized as follows: Section 2 reviews related literature and studies. Section 3 outlines our proposed methods. Section 4 describes the pilot study used to verify our approach effectiveness. Section 5 details the main study conducted to evaluate the proposed method. Finally, conclusions are presented in Section 6.

## 2. Related work

### 2.1 Effects of the anger of disinformation

Factors influencing sharing behavior include social media engagement rate [30], [31], opinion leadership [32], high information value [33], authenticity judgments [34], predispositions [35], [36], and emotions. Several studies highlight the strong influence of emotions [37], [38]. Emotions affect cognitive [39], decision-making [40], and behavioral [41] processes. For example, the more people rely on emotions rather than reason to judge the accuracy of fake news, the more likely they are to perceive it as accurate [8]. The influence of emotions varies based on the type of emotion. For instance, angry individuals are more likely to rely on simple cues (stereotypes, source expertise, and source trustworthiness) [19] and provide a lower percentage of correct answers compared to those experiencing sadness [20]. Content evoking awe, anger, or anxiety is more likely to be shared, whereas content that evokes sadness spread less [38]. Anger arises from the perception that an unfair or unjust act has been committed against oneself or one's group [42], [43]. Once triggered, anger colors perceptions, influencing decisions, and directs behavior, regardless of the source [44].

While anger has societal downsides, it can serve to speak truth to power [45]. However, on social media, anger is often associated with more harmful consequences than benefits [46]. It induces decision-making that hampers the resolution of social conflicts [46]. Disinformation exploits anger to provoke groups into pursuing harmful goals, such as ineffective or counterproductive actions (e.g., calls to boycott elections [7]). According to Steinert and Dennis, for many users, the negative emotions of social comparison do not help them in their pursuit of values but merely paralyze them, failing to promote the best actions that lead to the wellbeing of user [13]. Allowing reason to prevail over anger in social issues related to disinformation may help guide individuals and groups toward the right goals and best actions.

### 2.2 Strategies for reflection on anger

To become aware of emotions and make thoughtful decisions, emotional intensity should be balanced, not too high or low [47], [48]. Excessive emotional intensity leads to impulsive responses, while insufficient intensity results in numbness and difficulty engaging in thoughtful reflection. Depending on the situation, people regulate their emotions by adjusting their emotional arousal upward or downward, such as by expressing laughter, distress, or anger. Emotion regulation is defined as "the processes by which individuals influence which emotions they have, when they have them, and how they experience and express these emotions" [27]. Gross proposed an emotion-regulation process model comprising four modes of emotion generation: situation, attention, appraisal, and response. Each mode is associated with five emotion regulation strategies: situation selection, situation modification, attentional deployment, cognitive change, and response modulation [49]. Within these, subtypes exist for attentional deployment, cognitive change, and response modulation.

Attentional deployment strategies include distraction, concentration, and rumination [27]. Distraction focuses on the non-emotional aspects of a situation or redirect attention away from the immediate source of emotion [27]. Distractions encouraging individuals to think about positive, unrelated matters emotionally arousing stimuli or emotions have proven effective [50]. Concentration, on the other hand, involves focusing on the emotional aspects of a situation. When attention repeatedly returns to an emotion and its consequences, this process is referred to as rumination [49]. Both concentration and rumination can amplify anger by encouraging the re-experiencing of emotion-triggering memories [51]. Cognitive change strategies include reappraisal of emotional stimuli and perspective-taking [50]. Reappraisal involves mentally modifying how a situation is evaluated before a full emotional response is elicited [52]. For instance, reappraisal may include adopting a positive mindset by imagining favorable outcomes in potentially negative scenarios [50], [53]. Perspective-taking is the spontaneous effort to adopt the perspectives of other people and to see things from their point of view [54]. This ability helps anticipate others' behaviors and reactions, fostering smoother and more rewarding interpersonal interactions [55]. Suppression is a common strategy within response modulation [50]. Roberton et al. found that controlling the outward expression of anger does not necessarily reduce aggressive behavior. Instead, they suggested that the ability to control behavior when feeling angry is important [56]. They also recommended promoting the separation of emotions and behavior, noting that emotions can be tolerated without necessitating action [56]. Emotion regulation can also influence the emotional trajectories of others, a process referred to as extrinsic emotion regulation [49]. Perspective-taking and empathic responses have been shown to counter negative emotions effectively [57]. Empathic responses refer to efforts to engage with others in the response modulation mode of conveying understanding, validation, and care for their emotional experience [57].

## 3. Proposed digital nudge design

### 3.1 Implementation requirements

Digital nudges were displayed when users shared posts with high anger scores, based on the emotional rating of disinformation. This is because it is helpful for users to regain their emotions by addressing the strong anger associated with disinformation and to post or share based on careful deliberation. We designed the system to respect user's autonomy, allowing them to either follow or ignore the sharing behavior, while providing tips to remind users of the desired rational state through emotional regulation, encouraging them to deliberate. Crucial elements of successful emotion regulation are emotion differentiation and knowing how to flexibly regulate emotions [58]. Emotion differentiation refers to the accuracy with which people can identify and distinguish their emotions [59]. If an individual is unaware of their emotional response, it is unlikely they will regulate it effectively [58]. Webb et al. [60] suggested that the first task in regulating emotions is identifying the need for regulation. This need arises when there is a discrepancy between a person's current emotional state and their desired state, as defined by their emotional standards [60]. This discrepancy occurs when emotions are too strong, too frequent, last too long, or the wrong type of emotion for the situation [61]. This comparator function is triggered when attention is drawn to current emotions [28].

To design the digital nudge, we incorporated two new mechanisms: displaying specific emotional information and providing emotion regulation messages. First, emotional information was displayed, showing the types of emotions contained in the disinformation and their percentages. These nudges drew attention to emotional information and provided comparisons for reflecting on emotions. Kiskola et al. [62], [61] and Syrjämäki et al. [63] made users aware that the text contained emotional expression, but did not clearly indicate the emotion type. To improve this, it is necessary to provide emotional information that specifically indicates the emotion type. Second, we displayed emotion regulation messages that provided tips to encourage deliberation. These nudges reminded users of their rational state based on emotion regulation strategies and may have prevented emotional sharing. When users notice a discrepancy between emotional information and their emotional standards (e.g., if the emotion evoked was too strong or the wrong type of emotion was aroused [61]), they could reconsider sharing by using the message tips to encourage deliberation. Emotional information and emotion regulation messages were implemented through a friction intervention that displayed a pop-up window at the moment of sharing. Friction refers to a design that intentionally makes processes slower or more effortful [14], which can reduce the sharing of false information [22].

### 3.2 Displayed elements for digital nudges

The social media platform assumed to display the digital nudges is X. X had the highest ratio of disinformation discoverability, meaning the proportion of sensitive content identified as disinformation in a European Commission survey [64]. In a Japanese survey, X was also the medium through which users were most exposed to misinformation or misleading information [65]. To minimize the influence of novelty on the user, the design and response options displayed were similar to those introduced by X in June 2020 [23], [66] (Fig. 1). This feature encourages the user to read the article they are about to share and provides a button to choose between repost (share) or quote (add comment and share). Users can cancel the sharing by tapping outside the pop-up window.

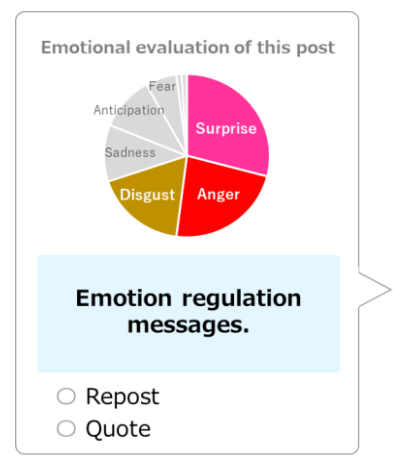

Figure 1 Digital nudge design to encourage deliberation by drawing attention to emotional information.

Emotional information is a graphical visualization of the type of emotion contained in the disinformation. This approach aligns with recommendations to prioritize visual techniques when building technology related to user-sharing behaviors [67]. Visualization is a form of intervention design used as an interface [68], and it is an effective and transparent way to provide information while offering an explanation [68]. Pandey et al. found that charts were more persuasive than textual or tabular information, except when participants held strong initial attitudes about the topic (e.g., skepticism or anchoring to core beliefs) [69]. Some studies have used color bars or pie charts to show the probability of misinformation authenticity, measuring their effect on users' trust in the information and their intent to share it [67], [70]. Amin et al. discussed that when users decided to share misinformation, the visual intervention caught their attention and led them to cancel the sharing process [67]. Regarding the visualization of emotional information, one study proposed a web application that displayed sentiment analysis results for articles on a news site as a sentiment score, using a color bar [71]. Emotion regulation studies have employed various methods to help individuals focus on or distract themselves from their emotions, with online environments attempting to support emotion regulation through user interface design [62], [61], [63]. Kiskola et al. proposed interface designs that made users aware of emotional information in the text when writing uncivil comments on online news articles [62]. This is based on the concept that by helping users perceive emotionally loaded elements, self-reflection and emotion regulation can be encouraged [61].

We presented emotion regulation messages as textual feedback to users along with emotional information. This approach is based on the idea that graphical visualization can be enhanced by providing accompanying text messages [72], [73]. Users prefer not only visual information but also practical and detailed feedback, such as suggestions for improvement or concrete modifications based on the provided information [73]. Infographics, which combine visuals and text to graphically display information, reduce cognitive load and increase users' ability to assess false news [74]. The emotion regulation messages employed various emotion regulation strategies, excluding concentration and rumination, which have been shown to increase anger in related studies. The messages included three distractions techniques [75], [76], two reappraisals [77], [78], [79], [80], two perspective-taking strategies [81], [82], [83], [84], and one empathic response [57]. Nine digital nudges were created: one displaying only emotional information, and another displaying both emotional information and emotion regulation messages (Table 1). Behavioral suppression was defined as the suppression of sharing behavior through friction.

Table 1 Proposed digital nudges (nine types)

| Emotion regulation | ID | Message overview |
|---|---|---|
| Pie chart only | A1 | No message. |
| Distraction | D1 | Imagine positive situations. |
| | D2 | Imagine compassion toward contributors. |
| | D3 | Imagine compassion toward followers. |
| Reappraisal | R1 | Connect situations to positive change. |
| | R2 | Change the interpretation of the emotional stimulus itself. |
| Perspective-taking | P1 | Incorporate self into others. |
| | P2 | Objective and analytical perspective. |
| Empathic response | E1 | Convey compassion toward yourself. |

## 4. Pilot study

We conducted a pilot study to identify digital nudges that were effective in decreasing the sharing of disinformation stimuli. This study, along with all subsequent experiments, was approved by the Research Ethics Committee prior to data collection.

### 4.1 Methodology
#### 4.1.1 Participants

Participants for the pilot study were recruited from a graduate school and the company to which the experimenter was affiliated. A total of 47 participants (30 men and 17 women; age = 10s-60s) provided responses. Some participants were aware of the aim of the pilot study. To minimize bias stemming from participants responding in line with the experimenter's desired outcome, a role-playing approach was employed. Participants were asked to adopt a third-person perspective, responding as if they were someone else. The pilot study was conducted using Google Forms. Participants were presented with an image of text-posting stimuli that appeared in the recommended content of X's home timeline. They were instructed to respond to questions as if they were a person of the same gender who had pressed the repost button out of strong anger.

#### 4.1.2 Stimuli

From the text-posting stimuli developed by Suzuki and Inaba [10] to mimic X's text posts (Table 2), we used two disinformation stimuli focused on the topic of gender conflict: disinformation targeting men and women. The stimuli consisted of approximately 140-character sentences, adhering to X's character limits for posts in Japanese-speaking countries. Two topics were set: gender and generational conflict have been identified as key inequality issues of interest in Japan.

Table 2 The text-posting stimuli by Suzuki & Inaba [10]

| Conditions | Social anxiety | Felt unjust |
|---|---|---|
| Disinformation (men/older) | strong | yes |
| Disinformation (women/younger) | strong | yes |
| Disinformation-control | weak | yes |
| True information-emotional | strong | no |
| True information-control | weak | no |

Note: Ten stimuli under five conditions each on the topics of gender and generational conflict.

The disinformation stimuli were modeled on the text and structure of the 3,266 disinformation cases from the 2016 U.S. election, as published by the House Permanent Select Committee on Intelligence (HPSCI) [85]. These past cases aimed to incite emotions by suggesting that one's associated identity was being unfairly treated by a conflicting group, thereby exploiting social anxiety. Research suggests that false rumors spread more effectively than truth-based posts when they contain a high number of other-condemning emotion words. Conversely, posts with more self-conscious emotion words are associated with a less viral spread [86]. The disinformation-control stimuli were designed to express unfairness toward conflicting group but incorporated more self-conscious emotions, which reduced their emotional impact and did not provoke strong emotions. The true information stimuli were summaries of news articles from traditional or web media in Japan. These stimuli avoided content that could fuel a sense of unfairness toward conflict groups. True information-emotional stimuli evoked participants' emotions by leveraging social anxiety. True information-control stimuli did not evoke emotional reaction, as they were crafted with weaker social anxiety elements.

### 4.1.3 Emotional information in digital nudges

The emotional information displayed in the digital nudges was based on data collected from 300 participants in an earlier experiment conducted by Suzuki and Inaba [10]. In the previous experiment, participants were exposed to disinformation stimuli created by the experimenter and were asked to identify the type of emotion aroused by the stimuli using Plutchik's Wheel of Emotions [87]: anticipation, joy, trust, fear, surprise, sadness, disgust, anger, or non-emotion. In the current study, the percentage of past responses indicating each emotion type was visualized using pie charts (Table 3). The three most frequent reported emotions were highlighted in their associated colors [88], while the rest were grayed out. The colors assigned to each emotion were bright pink for surprise, red for anger, dark yellow for disgust, and indigo for sadness (Fig. 2).

Table 3 Emotional information of disinformation stimuli

| Topics | Stimuli | 1st emotion | 2nd emotion | 3rd emotion |
|---|---|---|---|---|
| Gender conflict | Disinformation (men) | surprise (29.1%) | anger (23.0%) | disgust (17.9%) |
|  | Disinformation (women) | anger (32.6%) | sadness (28.9%) | surprise (14.4%) |
| Generational conflict | Disinformation (older) | anger (39.3%) | disgust (24.1%) | sadness (15.7%) |
|  | Disinformation (younger) | sadness (30.7%) | anger (30.3%) | surprise (17.3%) |

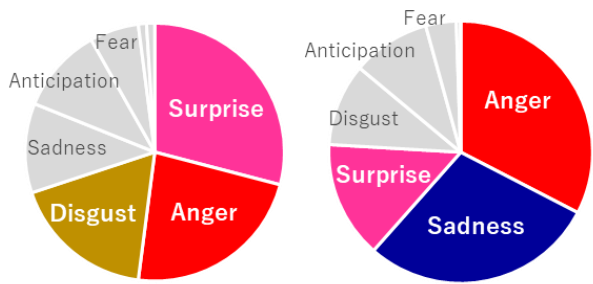

Figure 2 Graphical visualizations of the emotional information used in the digital nudges; The left figure shows disinformation (men) stimulus, while the right figure shows disinformation (women) stimulus.

### 4.1.4 Procedure

To help participants visualize strong anger someone of the same gender, men were presented with a disinformation stimulus addressing a social issue they perceived as unfair compared to women. Conversely, women were shown a disinformation stimulus that made them feel unfairly treated compared to men. Nine digital nudges were then presented randomly in ascending or descending order. For each digital nudge presentation, participants were asked to respond to two questions: change in sharing intentions (response options: repost/quote/cancel) and change in emotional intensity (5 = pre-nudge intensity, 0 = weak to 10 = strong).

### 4.2 Results

We evaluated the effectiveness of the proposed digital nudges in reducing the sharing of disinformation stimuli. For each nudge, we calculated the percentage of participants selecting each sharing intention response (repost/quote/cancel) (Table 4). The results showed that A1 had the highest percentage of participants choosing to repost, while P2 and D3 had the highest percentages of participants who chose to cancel. Chi-square tests were conducted for each digital nudge to analyze the number of respondents selecting each sharing intentions (repost/quote/cancel). P2 showed a significantly higher number of participants canceling compared to those reposting ($\chi^2(2) = 24.553, p < .01$).

We also examined changes in emotional intensity after nudging. Using pre-nudging emotional intensity as the baseline (5), median and quartile deviations were calculated for post-nudge emotional intensity (scale: 0-10) (Fig. 3). The results revealed that emotional intensity slightly strengthened after nudging conditions for E1 ($Md = 5$, $QD = 5$-6), and weakened for P2 ($Md = 4$, $QD = 2.0$-$5.0$) and D3 ($Md = 4$, $QD = 2.0$-$5.5$). Chi-square tests were conducted for each stimulus to assess whether the number of respondents with weaker or stronger emotional intensity after nudging differed significantly. The results indicated that P2 and D3 had significantly more respondents with weaker emotional intensity than stronger emotional intensity (P2: $\chi^2(1) = 18.778, p < .001$; D3: $\chi^2(1) = 4.568, p < .05$).

Table 4 Response tendencies of sharing intentions

| ID | Repost (%) | Quote (%) | Cancel (%) |
|---|---|---|---|
| A1 | 61.7 | 19.1 | 19.1 |
| D1 | 44.7 | 17.0 | 38.3 |
| D2 | 51.1 | 21.3 | 27.7 |
| D3 | 42.6 | 8.5 | 48.9 |
| R1 | 27.7 | 42.6 | 29.8 |
| R2 | 55.3 | 27.7 | 17.0 |
| P1 | 53.2 | 17.0 | 29.8 |
| P2 | 25.5 | 8.5 | 66.0 |
| E1 | 40.4 | 34.0 | 25.5 |

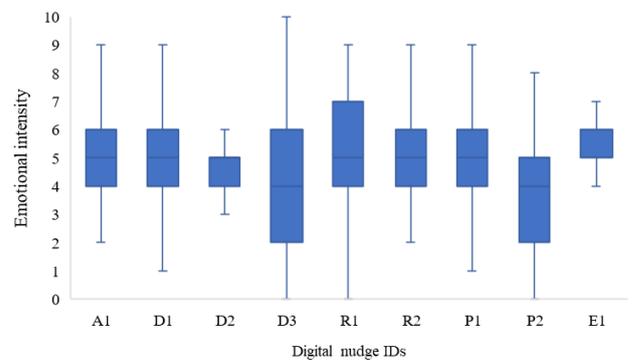

Figure 3 Post-nudge emotional intensity. Pre-nudge emotional intensity (baseline) = 5.

### 4.3 Discussion

The combination of emotional information and emotion regulation messages was more effective than emotional information alone in reducing the sharing of disinformation stimuli. Specifically, the combination of imagined compassion toward followers (D3) and objective, analytical perspective (P2) may encourage deliberation and counteract the strong angry-driven sharing of disinformation stimuli.

These findings align with those of Webb et al. [50], who identified effective methods for managing emotions based on the inter-process effect size of emotion regulation strategies. However, this study had several limitations. First, the responses in the pilot study relied on participants imagining a person with strong anger, meaning the intervention's direct effect on participants' actual sharing behavior was not verified. Second, the instructions emphasizing strong anger toward disinformation may have introduced biases to the result of the emotional intensity. Third, presenting different text-posting stimuli to men and women may have influenced the participants' responses. To address these limitations, further research is needed to determine whether the two types of digital nudges that encourage deliberation are effective in influencing actual sharing behaviors driven by participants' emotions in response to disinformation stimuli.

## 5. Main study

### 5.1 Hypotheses

The effects of two digital nudges (D3/P2) that encouraged participants to focus on strong anger and deliberation when deciding to share were evaluated by comparing them with existing digital nudges. The existing nudges used for comparison were modeled after the features introduced by X in October 2020, which encouraged quoting over reposting [24]. This feature introduced friction, giving users an extra moment to consider why and what they were adding to their conversations. Consequently, X reported a 23% reduction in reposts [89]. To compare the effects of the three types of digital nudges, we measured sharing intentions, the type and intensity of emotion, and authenticity before and after the nudges were applied. Our proposed distraction and perspective-taking nudges were designed based on the prediction that emotional information and emotion regulation messages would direct users' attention to their emotions and encourage more deliberate behavior. While existing nudges effectively reduce sharing through friction-based mechanism, distraction and perspective-taking nudges incorporate emotion regulation strategies to promote deliberation. Therefore, we hypothesized that distraction and perspective-taking nudges would be more effective in leading participants to reconsider their sharing than existing nudges with only friction, and that they would reduce sharing driven by strong anger associated with disinformation. Our proposed hypotheses are as follows:

**H1.** Distraction, perspective-taking, and existing nudges decreased the sharing of disinformation more than pre-nudges.

**H2.** Distraction and perspective-taking nudges result in less sharing of disinformation compared to existing nudges.

### 5.2 Methodology

#### 5.2.1 Participants

The participants included 400 individuals (200 men and 200 women; $M$ age = 44.50, $SD$ = 13.58) living in Japan, who were recruited in January 2024 as online survey monitors through a web-based research company. Participants were compensated with reward program incentives. All participants met the following criteria: they used X at least three times per week and had previously reposted content from the recommended timeline that appears upon logging into X. The study was designed based on the X's interface, making it necessary for participants to have a regular familiarity with the platform and to understand terms such as button names and associated actions. Informed consent was obtained from all participants prior to the study. After excluding responses based on a lie scale, the final valid sample consisted of 400 participants.

#### 5.2.2 Stimuli

Ten stimuli (Table 2) derived from Suzuki & Inaba [10] were used in the study. The types and intensities of emotions recognized by participants for each text-posting stimulus were measured, serving as a baseline for comparison with responses to disinformation stimuli.

#### 5.2.3 Procedure

The study was conducted using a questionnaire administered by a web-based research company. Participants were informed that the study was for academic purposes and that all posts presented during the experiment was fictional and created by the experimenter. Participants were shown an image of text-posting stimuli appearing in the recommended content of X's home timeline and were asked to respond to questions regarding their sharing intentions and the emotions they recognized.

Participants were randomly assigned to one of the two topics. Unlike the pilot study, five text-posting stimuli were presented randomly in one of two patterns, balanced across genders and ages. The first pattern presented the stimuli in the following order: true information-emotional, disinformation (men/older), true information-control, disinformation (women/younger), and disinformation-control. The second pattern presented the stimuli in reverse order. After the disinformation stimuli, participants were randomly exposed to the following three types of digital nudges: Distraction nudges that displayed emotional information and the message, "Imagine followers who always smile when they see your posts," perspective-taking nudges that displayed emotional information and the message, "This post may be intended to provoke emotional responses from the audience," and existing nudes that displayed only the message "Add comment."

After each text-posting stimulus or digital nudge, participants were asked to respond to four questions. The questions were:

1) Sharing intentions. Following the presentation of the text-posting stimuli, participants were asked to respond to the question before nudging, "Would you repost this post if it appeared in your recommended timeline?" (Response options: repost/not repost). Then, after nudging participants were asked to respond to the question, "Which button do you think you would click when you see the pop-up window displayed?" (Response options:

repost/quote/cancel).
2) Type of emotion. Participants were asked to select the closest emotion they recognized from Plutchik's Wheel of Emotions [87]: anticipation, joy, trust, fear, surprise, sadness, disgust, anger, or non-emotion.
3) Intensity of emotion. Before nudging, if any emotion was recognized, the participants were asked to rate its intensity on a scale of 1 (weak) to 10 (strong). After nudging, participants were asked to rate the intensity of the emotion on a scale of 1 (weaker) to 10 (stronger).
4) Authenticity. Participants were asked, "Do you believe that the text-posting stimulus was a real event?" (Response options: believe/disbelieve).

At the end of the experiment, participants were asked an additional question, to evaluate the ethical considerations of the study, by rating their "post-feelings" on a scale from 1 (positive) to 5 (negative).

## 5.3 Results
### 5.3.1 General response tendencies

Before examining the hypotheses, we analyzed general response tendencies. The percentage of participants who indicated they would share each text-posting stimulus was calculated (Table 5). For the gender conflict text-posting stimuli, the participants' mean sharing rate was 6.8% ($SD$ = 2.2), with the sharing rates for disinformation stimuli hovering around the mean. For the generational conflict text-posting stimuli, the mean sharing rate was higher at 14.9% ($SD$ = 3.3). The disinformation stimulus targeting younger participants had the highest sharing rate, while the disinformation stimulus targeting older participants was close to the mean. Overall, all disinformation stimuli demonstrated average or above average sharing rates.

We also analyze the intensity and type of emotions participants associated with each text-posting stimulus. Emotional stimuli, including disinformation and true information-emotional stimuli, were grouped together, while disinformation-control and true information-control stimuli were grouped as control stimuli. Wilcoxon signed-rank tests were performed, using emotional and control stimuli as independent variables, and emotional intensity (measured on a scale of 0-10) as the dependent variable. The results showed significant differences in emotional intensity between disinformation and disinformation-control stimuli, as well as between true information-emotional and true information-control stimuli, for both topics (all $p$ < .005). For comparison, we ranked the intensities of the emotions and calculated their average rankings. Disinformation stimuli (219.0-221.2) were ranked higher than disinformation-control stimuli (179.8-180.9), and true information-emotional stimuli (212.9-222.3) were ranked higher than true information-control stimuli (178.7-188.1). These results confirm that participants recognized stronger emotions for disinformation and true information-emotional stimuli. In addition, we calculated the percentage of emotional responses participants reported for each text-posting stimulus (Fig. 4). For disinformation stimuli, participants most frequently recognized emotions such as disgust, sadness, and anger. In contrast, true information stimuli were associated primarily with sadness and anticipation.

Table 5 The percentage of shared for each stimulus

| Topics | Text-posting stimuli | Share (%) |
|---|---|---|
| Gender conflict ($n$=200) | Disinformation (men) | 8.0 |
| | Disinformation (women) | 6.0 |
| | Disinformation-control | 4.0 |
| | True information-emotional | 5.5 |
| | True information-control | 10.5 |
| Generational conflict ($n$=200) | Disinformation (older) | 13.0 |
| | Disinformation (younger) | 21.2 |
| | Disinformation-control | 13.0 |
| | True information-emotional | 14.9 |
| | True information-control | 12.4 |

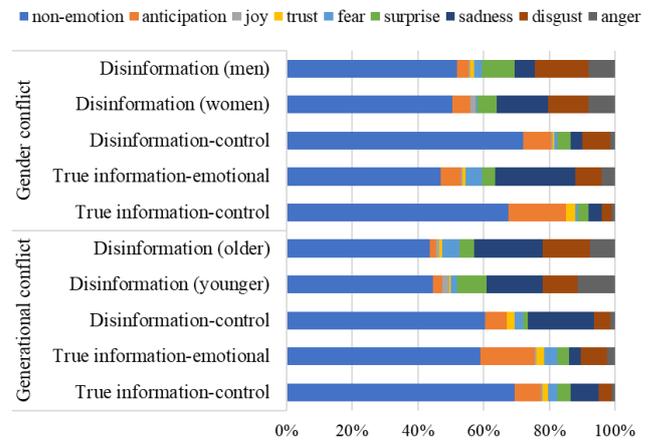

Figure 4 Emotion types for each stimulus.

We also examined the percentage of belief and shared responses for each text-posting stimulus. Since true information stimuli were summaries of media news, we investigated the potential influence of familiarity on sharing intentions. The belief responses percentages for true information stimuli ranged from 40-52%, while for disinformation stimuli, the range was 40-61%. Chi-square tests were conducted for each topic, with text-posting stimuli as the independent variable and the number of believers as the dependent variable. For gender conflict, no significant differences were found in the percentage of respondents who believed the stimuli. However, for generational conflict, the number of believers was significantly higher for the disinformation-control stimuli than for the true information-emotional stimuli ($p$ < .01). These findings suggest that participants were generally unfamiliar with the true information stimuli and that the influence of familiarity on sharing intentions was minimal.

From an ethical perspective, we evaluated participants' emotional states (measured on a scale from 1 = positive to 5 = negative) after the experiment. The median and quartile deviations of the participants' responses were calculated, and the median was neutral at 3 ($QD$ = 3-4), indicating no extreme negative bias in participants' emotions after the study.

### 5.3.2 Effects of friction on sharing

To examine changes in sharing intentions before and after nudging, we focused on participants who shared disinformation stimuli prior to nudging. After each digital nudge, the percentage of participants who continued to share (repost or quote) was 66.3% for distraction nudges, 69.8% for perspective-taking nudges, and 77.9% for existing nudges.

To evaluate H1, we conducted Cochran's Q tests with pre-nudge and post-nudge conditions (for all three digital nudges) as independent variables and sharing intention responses as the dependent variable. For analysis, pre-nudge-sharing intentions were converted to numerical data: repost = 1 and not repost = 0. Similarly, post-nudge-sharing intentions were also converted with repost or quote = 1 and cancel = 0. The results revealed a significant difference in sharing intentions between before and after nudging ($Q = 50.9$, $df = 3$, $p < .001$). Multiple comparisons using the Bonferroni correction showed that all digital nudges significantly decreased sharing intentions compared to before nudges conditions ($p < .001$).

### 5.3.3 Effects of deliberation on sharing

To evaluate H2, we conducted a Cochran's Q test with the three digital nudges as independent variables and sharing intention responses as the dependent variable. Sharing intentions were converted into numerical data: repost or quote = 1 and cancel = 0. The results indicated a significant difference in sharing intentions among the digital nudges ($Q = 6.87$, $df = 2$, $p < .05$). Multiple comparisons using the Bonferroni correction revealed that distraction nudges were shared significantly less frequently than existing nudges ($p < .05$). However, no significant difference was observed between perspective-taking and existing nudges.

To examine the effect of each digital nudge on specific disinformation stimuli, Cochran's Q test were conducted for each text-posting stimulus. The results showed that distraction nudges significantly reduced the sharing of all disinformation stimuli ($p < .05$). Perspective-taking nudges significantly decreased sharing of two disinformation stimuli (targeting older and younger participants), while existing nudges significantly reduced sharing of the younger-focused of disinformation stimulus (all $p < .05$). These findings suggest that all digital nudges significantly reduced sharing, though the extent of the effect varied depending on the content and messages presented to users (Fig. 5).

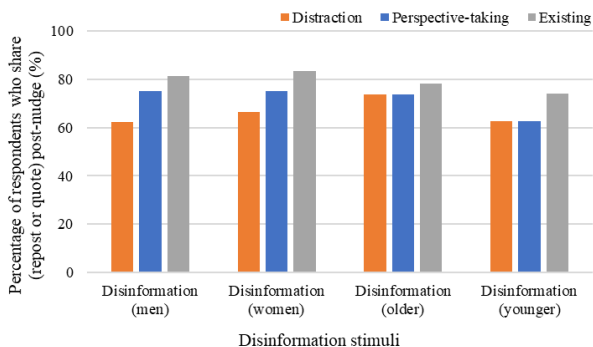

Figure 5 Percentage of post-nudge sharing (repost or quote) for each disinformation stimulus among participants who initially responded that they would share the disinformation stimuli.

### 5.3.4 Effects of deliberation through emotion regulation

To compare the effects of deliberation through emotion regulation, we calculated the median and quartile deviations for emotional intensity before and after nudging among participants who initially responded that they would share disinformation stimuli. Across all digital nudges, emotional intensity was lower after nudging ($Md = 6$, $QD = 5$-$8$) than before nudging ($Md = 7$, $QD = 6$-$8.75$). A Friedman test was conducted with before nudges and the three digital nudges as independent variables and emotional intensity responses as the dependent variable. The result of the test showed a significant difference in emotional intensity ($\chi^2(3) = 22.617$, $p < .001$). Multiple comparisons using the Bonferroni correction indicated that distraction and existing nudges significantly reduced emotional intensity compared to before nudges (all $p < .05$).

Among participants whose emotional intensity was reduced after the nudges, the percentage who canceled sharing disinformation was higher for distraction nudges (41.5%), perspective-taking nudges (39.4%), and existing nudges (19.5%) (Fig. 6). Chi-square tests were conducted for each digital nudge, using participants' sharing intentions (repost or quote/cancel) after the nudge as the independent variable and the number of respondents as the dependent variable. The results showed no significant difference between shared and canceled responses for distraction and perspective-taking nudges among participants whose emotions weakened after nudging. However, for existing nudges, sharing was significantly more likely than canceling among participants whose emotions weakened after the nudge ($\chi^2(2) = 5.308$, $p < .05$).

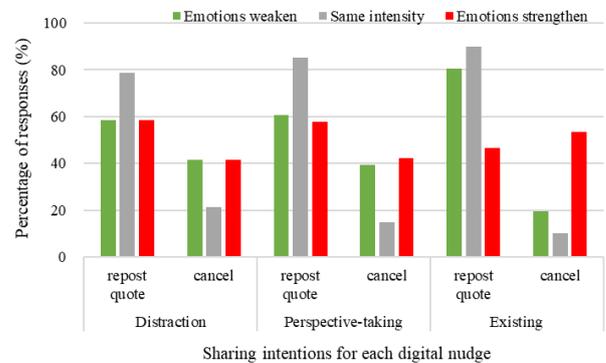

Figure 6 Response tendencies of sharing intentions based on changes in emotional intensity after nudging among participants who responded that they would share the disinformation stimuli.

Distraction nudges also demonstrated the potential to positively shift emotions responses to disinformation stimuli. Among participants who initially recognized and shared negative emotions such as fear, sadness, disgust, and anger in response to disinformation stimuli, a small percentage experienced a shift to positive emotions such as anticipation, joy, or trust after nudging. This change was observed in 1.5% of participants for distraction nudges, compared to 0.8% for both perspective-taking and existing nudges.

### 5.4 Discussion

To the best of our knowledge, this study is the first to focus on the emotions of social media users as an approach in reducing the sharing of online disinformation. This study evaluated whether digital nudges that encourage deliberation by drawing attention to emotional information could reduce sharing driven by strong anger associated with online disinformation. H1 was supported, as all digital nudges were effective in decreasing the sharing of disinformation stimuli, with many participants recognizing anger. Comparisons on the effects of different digital nudges, partially supported H2, as distraction nudges were more effective than existing nudges at reducing the sharing of disinformation stimuli.

The distraction nudges used emotion regulation messages designed to remind compassion toward followers. Participants were encouraged to imagine "followers who always smile when they see your posts," creating a discrepancy with the emotional information [61] in the disinformation stimuli, which included anger, sadness, disgust, and surprise (see Table 3). This discrepancy may have triggered the need for emotion regulation [60], prompting participants to deliberate using emotion regulation strategies. Furthermore, the sharing of negative emotions may have been reduced because participants were concerned about making their followers feel uncomfortable.

In contrast, perspective-taking nudges were designed to encourage participants to consider the intent behind the post by providing information that prompted an objective and analytical perspective. By directing the participants' attention to the intent of the disinformation stimuli, these nudges may have prevented them from focusing on their emotions, making it harder to detect discrepancies [61]. As a result, the need for emotion regulation [60] was not activated, and deliberation using emotion regulation was not encouraged. Perspective-taking nudges also do not require participants to consider other's feelings when deciding whether to share. Based on the observed differences between the effects of distraction and perspective-taking nudges, increasing users' awareness of others and encouraging them to imagine the potential impact of sharing disinformation may be a particularly effective strategy for reducing sharing.

Although existing nudges reduce the sharing of disinformation stimuli, their effect was smaller than that of distraction nudges. While the reduction in sharing observed for all digital nudges may be partly attributed to the friction effect [22], distraction nudges were shared significantly less frequently than existing nudges. This suggest that the additional effects of distraction nudges stem from the inclusion of emotional information and emotion regulation messages, beyond the friction effect alone. Moreover, existing nudges appeared to continue sharing of disinformation stimuli primarily among participants whose emotional responses weakened after exposure to the nudge. This indicates that the cognitive processes underlying the effects of different digital nudges may vary. Distraction nudges may lead participants to reconsider sharing through deliberate cognitive processes using emotion regulation [26]. In contrast, existing nudges may rely on intuitive cognitive processes [19], [20] which are less likely to involve deliberation.

### 5.4.1 Limitations and future directions

This study has three limitations. First, the emotional information used in this study was based on previous evaluation results [10], whereas real-time sentiment analysis would be necessary for actual implementation. Sentiment analysis can be conducted either mechanically or by having users rate sentiments, as in the case of X's community notes [90]. Based on these evaluation results, an implementation method could involve presenting digital nudges when a specific sentiment type exceeds a predefined threshold value (e.g., [91]).

Second, the influence of digital nudges on emotionally evocative true information was not examined. Digital nudges may also be applied to true information by targeting the emotional content within a post. True news often elicits emotions such as anticipation, joy, trust, or sadness [2]. In the present study, true information stimuli typically recognized anticipation or sadness in participants. For distraction nudges, emotions like anticipation, joy, or trust may not create a need for emotion regulation due to their emotional alignment with the imagined smiling followers. Additionally, sadness is less likely to prompt sharing [10], [38], suggesting that the influence of distraction nudges on sharing true information may be minimal.

Third, our findings were based on results obtained in controlled experimental settings. While the main study was designed to eliminate suggestibility effects (a concern identified during the pilot study), the results were obtained within a laboratory. Therefore, the possibility that the experimental design influenced the results cannot be completely ruled out. The percentage of existing nudges, which mimicked the features of X and reduced sharing, was similar to the measured value [89]. However, it is important to test whether the effects of distraction nudges can be replicated in real-life scenarios.

### 6. Conclusion

In this study, we clarified the digital nudges that effectively reduce the sharing of user behaviors driven by strong anger associated with online disinformation. We designed digital nudges that encourage deliberation by drawing attention to emotional information as a novel approach. When evaluating their effectiveness, we found that distraction nudges, which combined emotional information with emotion regulation messages (specifically imagining compassion toward followers), were particularly effective. These nudges likely increased users' awareness of others and helped them consider the impact of their post on others, thereby effectively reducing the likelihood of sharing. These findings suggest that digital nudges can offer effective alternatives to traditional authenticity judgments in preventing the spread of online disinformation.

As a future extension, we aim to verify whether the proposed digital nudges can help users effectively utilize the knowledge and competencies acquired through educational interventions. By validating fact-checked misinformation, we plan to examine whether these nudges not only reduce the sharing of disinformation but also prevent the formation of false beliefs.

# Reference

[1] Pennycook, G., Epstein, Z., Mosleh, M., et al.: Shifting attention to accuracy can reduce misinformation online, Nature, Vol.592, pp.590-595 (2021).

[2] Vosoughi, S., Roy, D. and Aral, S.: The spread of true and false news online, Science, Vol.359, pp.1146-1151 (2018).

[3] European Commission: COM (2018) 236 final, EUR-Lex Access to European Union Law (2018).

[4] High Level Expert Group: A multi-dimensional approach to disinformation, European Commission (2018).

[5] Rid, T.: Active Measures: The Secret History of Disinformation and Political Warfare, p.513, Farrar Straus & Giroux (2020).

[6] Claire, W. and Hossein, D.: Information disorder: Toward an interdisciplinary framework for research and policymaking, Council of Europe (2017).

[7] DiResta, R., Shaffer, K., Ruppel, B., et al.: The tactics & tropes of the internet research agency, DigitalCommons@University of Nebraska - Lincoln (2019).

[8] Martel, C., Pennycook, G. and Rand, D. G.: Reliance on emotion promotes belief in fake News, Cognitive Research: Principles and Implications, Vol.5, Article No.47 (2020).

[9] Lewandowsky, S., Ecker, U. K. H., Seifert, C. M., et al.: Misinformation and its correction: Continued influence and successful debiasing, Psychological Science in the Public Interest, Vol.13, pp.106-131 (2012).

[10] Suzuki, H. N. and Inaba, M.: Psychological study on judgment and sharing of online disinformation, *Proc. IEEE 47th Annual Computers, Software, and Applications Conference (COMPSAC)*, pp.1558-1563 (2023).

[11] Fan, R., Xu, K. and Zhao, J.: Higher contagion and weaker ties mean anger spreads faster than joy in social media, arXiv:1608.03656 (2016).

[12] Nabi, R. L.: Exploring the Framing Effects of Emotion: Do Discrete Emotions Differentially Influence Information Accessibility, Information Seeking, and Policy Preference?, Communication Research, Vol.30, pp.224-247 (2003).

[13] Steinert, S. and Dennis, M. J.: Emotions and Digital Well-Being: on Social Media's Emotional Affordances, Philosophy & Technology, Vol.35, Article No.36 (2022).

[14] Kozyreva, A., Lorenz-Spreen, P., Herzog, S. M., et al.: Toolbox of individual-level interventions against online misinformation, Nature Human Behaviour, pp.1044-1052 (2024).

[15] Featherstone, J. D. and Zhang, J.: Feeling angry: the effects of vaccine misinformation and refutational messages on negative emotions and vaccination attitude, Journal of Health Communication, Vol.25, pp.692-702 (2020).

[16] Evans, J. S. B. and Stanovich, K. E.: Dual-process theories of higher cognition advancing the debate, Perspectives on Psychological Science, Vol.8, pp.223-241 (2013).

[17] Kahneman, D.: Thinking, fast and slow, pp.499, New York: Farrar, Straus and Giroux (2011).

[18] Bago, B., Rand, D. G. and Pennycook, G.: Fake news, fast and slow: Deliberation reduces belief in false (but not true) news headlines, Journal of Experimental Psychology: General, Vol.149, No.8, pp.1608-1613 (2020).

[19] Bodenhausen, G. V., Sheppard, L. A. and Kramer, G. P.: Negative affect and social judgment: The differential impact of anger and sadness, European Journal of Social Psychology, Vol.24, pp.45-62 (1994).

[20] Park, H. and Daibo, I.: Effects of anger and sadness on detecting deception, Japanese journal of applied psychology, Vol.40, pp.1-10 (2014).

[21] Weinmann, M., Schneider, C. and Brocke, vom. J.: Digital Nudging, Business & Information Systems Engineering, Vol.58, pp.433-436 (2016).

[22] Fazio, L.: Pausing to consider why a headline is true or false can help reduce the sharing of false news, Harvard Kennedy School Misinformation Review, Vol.1 (2020).

[23] X Support [@Support]: June 11, 2020, X (online), available from <https://twitter.com/Support/status/1270783537667551233?s=20> (accessed on 2024-12-23).

[24] Gadde, V. and Beykpour, K.: Additional steps we're taking ahead of the 2020 US Election, X Blog (online), available from <https://blog.twitter.com/en_us/topics/company/2020/2020-election-changes> (accessed 2024-12-23).

[25] Levy, N.: Nudges in a post-truth world, Journal of Medical Ethics, Vol.43, pp.495-500 (2017).

[26] Muradova, L.: Seeing the Other Side? Perspective-Taking and Reflective Political Judgements in Interpersonal Deliberation, Political Studies, Vol.69, pp.644-664 (2021).

[27] Gross, J. J.: The emerging field of emotion regulation: An integrative review, Review of General Psychology, Vol.2, pp.271-299 (1998).

[28] Tamir, M.: Effortful Emotion Regulation as a Unique Form of Cybernetic Control, Perspectives on Psychological Science, Vol.16, pp.94-117 (2021).

[29] Stains Jr., R. R. and Sarrouf, J.: Hard to say, hard to hear, heart to heart: Inviting and harnessing strong emotions in dialogue for deliberation, Journal of Deliberative Democracy Vol.18, Issue 2 (2022).

[30] Stavrositu, C. D. and Kim, J.: Social media metrics: Third-person perceptions of health information, Computers in Human Behavior, Vol.35, pp.61-67 (2014).

[31] Chung, M., Munno, G. J. and Moritz, B.: Triggering participation: Exploring the effects of third person and hostile media perceptions on online participation, Computers in Human Behavior, Vol.53, pp.452-461 (2015).

[32] Ma, L., Lee, C. S. and Goh, D. H.: That's news to me: The influence of perceived Gratifications and personal experience on news sharing in social media, *Proc. 11th annual international ACM/IEEE joint conference on Digital libraries*, pp.141-144 (2011).

[33] Rudat, A., Buder, J. and Hesse, F. W.: Audience design in Twitter: Retweeting behavior between informational value and followers' interests, Computers in Human Behavior, Vol.35, pp.132-139 (2014).

[34] Jahanbakhsh, F., Zhang, A. X., Berinsky, A. J., et al.: Exploring Lightweight Interventions at Posting Time to Reduce the Sharing of Misinformation on Social Media, *Proc. ACM on Human-Computer Interaction*, Vol.5, Article No.18, pp.1-42 (2021).

[35] Conover, M. D., Ratkiewicz, J., Francisco, M., et al.: Political polarization on twitter, *Proc. International AAAI Conference on Web and social media*, Vol.5, No.1, pp.89-96 (2011).

[36] Fox, J., Cruz, C. and Lee, J. Y.: Perpetuating online sexism offline: Anonymity, interactivity, and the effects of sexist hashtags on social media, Computers in Human Behavior, Vol.52, pp.436-442 (2015).

[37] Ibrahim, A., Ye, J. and Hoffner, C.: Diffusion of news of the Shuttle Columbia Disaster: The role of emotional responses and motives for interpersonal communication, Communication Research Reports, Vol.25, pp.91-101 (2008).

[38] Berger, J. and Milkman, K. L.: What makes online content viral?, Journal of Marketing Research, Vol.49, pp.192-205 (2012).

[39] Bower, G. H.: Mood and memory, American Psychologist, Vol.36, No.2, pp.129-148 (1981).

[40] Lerner, J. S., Li, Y., Valdesolo, P., et al.: Emotion and Decision Making, Annual Review of Psychology, Vol.66, pp.799-823 (2015).

[41] Bless, H. and Fiedler, K.: Mood and the regulation of information processing and behavior, In Forgas, J. P. (Ed.), Hearts and Minds: Affective Influences on Social Cognition and Behavior, pp.65-84,

New York: Psychology Press (2006).

[42] Averill, J.: Anger and Aggression: An Essay on Emotion, pp.402, NY: Springer (1982).

[43] Shuman, E., Halperin, E. and Reifen-Tagar, M.: Anger as a catalyst for change? Incremental beliefs and anger's constructive effects in conflict, Group Processes & Intergroup Relations, Vol.21, pp.1092-1106 (2018).

[44] Lerner, J. S. and Tiedens, L. Z.: Portrait of The Angry Decision Maker: How Appraisal Tendencies Shape Anger's Influence on Cognition, Journal of Behavioral Decision Making, Vol.19, pp.115-137 (2006).

[45] Elsayed, Y. and Hollingshead, A. B.: Humor reduces online incivility, Journal of Computer-Mediated Communication, Vol.27, zmac005 (2022).

[46] Brady, W. J. and Crockett, M. J.: How effective is online outrage? Trends in Cognitive Sciences, Vol.23, pp.79-80 (2019).

[47] Siegel, D. J.: The developing mind: Toward a neurobiology of interpersonal experience, pp.394, The Guilford Press (1999).

[48] NHS Fife Psychology Department: Emotion Regulation: Managing Emotions, NHS Fife (online), available from <https://www.moodcafe.co.uk/media/fselmngo/er_handout_final_16_june_2016.pdf> (accessed 2024-12-23).

[49] Gross, J. J. and Thompson, R. A.: Emotion Regulation: Conceptual Foundations, In Gross, J. J. (Ed.), Handbook of emotion regulation, pp.3-24, The Guilford Press (2007).

[50] Webb, T. L., Miles, E. and Sheeran, P.: Dealing with feeling: A meta-analysis of the effectiveness of strategies derived from the process model of emotion regulation, Psychological Bulletin, Vol.138, pp.775-808 (2012).

[51] Rusting, C. L. and Nolen-Hoeksema, S.: Regulating responses to anger: Effects of rumination and distraction on angry mood, Journal of Personality and Social Psychology, Vol.74, pp.790-803 (1998).

[52] Denson, T. F., Moulds, M. L. and Grisham, J. R.: The effects of analytical rumination, reappraisal, and distraction on anger experience, Behavior Therapy, Vol.43, pp.355-364 (2012).

[53] Richards, J. M., Butler, E. A. and Gross, J.J.: Emotion Regulation in Romantic Relationships: The Cognitive Consequences of Concealing Feelings, Journal of Social and Personal Relationships, Vol.20, pp.599-620 (2003).

[54] Davis, M. H.: A Multidimensional Approach to Individual Differences in Empathy, JSAS Catalog of Selected Documents in Psychology, Vol.10, pp.85 (1980).

[55] Davis, M. H.: Measuring individual differences in empathy: Evidence for a multidimensional approach, Journal of Personality and Social Psychology, Vol.44, pp.113-126 (1983).

[56] Roberton, T., Daffern, M. and Bucks, R. S.: Beyond anger control: Difficulty attending to emotions also predicts aggression in offenders, Psychology of Violence, Vol.5, pp.74-83 (2015).

[57] Nozaki, Y. and Mikolajczak, M.: Effectiveness of extrinsic emotion regulation strategies in text-based online communication, Emotion, Vol.23, pp.1714-1725 (2023).

[58] Barrett, L. F. and Gross, J. J.: Emotional intelligence: A process model of emotion representation and regulation, in Emotions: Current issues and future directions, Mayne, T. J. and Bonanno, G. A. (Eds.), pp.286-310, The Guilford Press (2001).

[59] Lennarz, H. K., Lichtwarck-Aschoff, A., Timmerman, M. E., et al.: Emotion differentiation and its relation with emotional well-being in adolescents, Cognition and Emotion, Vol.32, pp.651-657 (2018).

[60] Webb, T. L., Gallo, I. S., Miles, E., et al.: Effective regulation of affect: An action control perspective on emotion regulation, European Review of Social Psychology, Vol.23, pp.143-186 (2012).

[61] Kiskola, J., Olsson, T., Syrjämäki, A. H., et al.: Online Survey on Novel Designs for Supporting Self-Reflection and Emotion Regulation in Online News Commenting, Proc. 25th International Academic Mindtrek Conference, pp.278-312 (2022).

[62] Kiskola, J., Olsson, T., Väätäjä, H., et al.: Applying critical voice in design of user interfaces for supporting self-reflection and emotion regulation in online news commenting, Proc. 2021 CHI Conference on Human Factors in Computing Systems, Article No.88, pp.1-13 (2021).

[63] Syrjämäki, A. H., Ilves, M., Kiskola, J., et al.: Facilitating Implicit Emotion Regulation in Online News Commenting—An Experimental Vignette Study, Interacting with Computers, Vol.34, pp.129-136 (2022).

[64] TrustLab: Code of Practice on Disinformation, TrustLab (online), available from <https://disinfocode.eu/wp-content/uploads/2023/09/code-of-practice-on-disinformation-september-22-2023.pdf> (accessed 2024-12-23).

[65] Ministry of Internal Affairs and Communications, Japan: Report on Information Circulation Survey on New Coronavirus Infections, MIC (online), available from 〈https://www.soumu.go.jp//000693280.pdf〉 (accessed 2024-12-23).

[66] Vincent, J.: Twitter is bringing its 'read before you retweet' prompt to all users, The Verge (online), available from 〈https://www.theverge.com/2020/9/25/21455635/twitter-read-before-you-tweet-article-prompt-rolling-out-globally-soon〉 (accessed 2024-12-23).

[67] Amin, Z., Ali, N. M. and Smeaton, A. F.: Visual Selective Attention System to Intervene User Attention in Sharing COVID-19 Misinformation, International Journal of Advanced Computer Science and Applications, Vol.12, pp.36-41 (2021).

[68] Hartwig, K., Doell, F. and Reuter, C.: The Landscape of User-centered Misinformation Interventions – A Systematic Literature Review, ACM Computing Surveys, Vol.56, Article No.292, pp.1-36 (2024).

[69] Pandey, A. V., Manivannan, A., Nov, O., et al.: The Persuasive Power of Data Visualization, IEEE Transactions on Visualization and Computer Graphics, Vol.20, pp.2211-2220 (2014).

[70] Ayoub, J., Yang, X. J. and Zhou, F.: Combat COVID-19 infodemic using explainable natural language processing models, Information Processing & Management, Vol.58, Article No.102569 (2021).

[71] Kim, P., Fan, Z., Fernando, L., et al.: Controversy Score Calculation for News Articles, Proc. First International Conference on Transdisciplinary AI (TransAI), pp.56-63 (2019).

[72] Ur, B., Kelley, P. G., Komanduri, S., et al.: How does your password measure up? The effect of strength meters on password creation, Proc. USENIX Security Symposium, pp.1-16 (2012).

[73] Ur, B., Alfieri, F., Aunget, M., et al.: Design and evaluation of a data-driven password meter, Proc. ACM Conference on Human Factors in Computing Systems, pp.3775-3786 (2017).

[74] Domgaard, S. and Park, M.: Combating misinformation: The effects of infographics in verifying false vaccine news, Health Education Journal, Vol.80, pp.974-986 (2021).

[75] Watson, C. and Purdon, C.: Attention training in the reduction and reappraisal of intrusive thoughts, Behavioural and Cognitive Psychotherapy, Vol.36, pp.61-70 (2007).

[76] Hutcherson, C. A., Seppala, E. M. and Gross, J. J.: Loving-kindness meditation increases social connectedness, Emotion, Vol.8, No.5, pp.720-724 (2008).

[77] Ochsner, K. N., Ray, R. D., Cooper, J. C., et al.: For better or for worse: Neural systems supporting the cognitive down- and up-regulation of negative emotion, NeuroImage, Vol.23, pp.483-499 (2004).

[78] McRae, K., Ochsner, K. N., Mauss, I. B., et al.: Gender differences in emotion regulation: An fMRI study of cognitive reappraisal, Group Processes & Intergroup Relations, Vol.11, pp.143-162 (2008).

[79] McRae, K., Hughes, B., Chopra, S., et al.: The neural bases of distraction and reappraisal, Journal of Cognitive Neuroscience, Vol.22, pp.248-262 (2010).

[80] Kalisch, R., Wiech, K., Critchleyet, H. D., et al.: Anxiety reduction through detachment: Subjective, physiological, and neural effects, Journal of Cognitive Neuroscience, Vol.17, pp.874-883 (2005).

[81] Ray, R. D., Wilhelm, F. H. and Gross, J. J.: All in the Mind's Eye Anger Rumination and Reappraisal, Psychological Science, Vol.94, pp.133-145 (2008).

[82] Galinsky, A. D. and Moskowitz, G. B.: Perspective-taking: Decreasing stereotype expression, stereotype accessibility, and in-group favoritism, Journal of Personality and Social Psychology, Vol.78, pp.708-724 (2000).

[83] Sheppes, G. and Meiran, N.: Better late than never? On the dynamics of online regulation of sadness using distraction and cognitive reappraisal, Personality and Social Psychology Bulletin, Vol.33, pp.1518-1532 (2007).

[84] Halperin, E., Pliskin, R., Saguy, T., et al.: Emotion regulation and the cultivation of political tolerance: Searching for a new track for intervention, The Journal of Conflict Resolution, Vol.58, pp.1110-1138 (2014).

[85] House Permanent Select Committee on Intelligence (HPSCI), US: Social Media Advertisements, Permanent Select Committee on Intelligence Democrats (2017).

[86] Solovev, K. and Pröllochs, N.: Moral Emotions Shape the Virality of COVID-19 Misinformation on social media, *Proc. ACM Web Conference 2022*, pp.3706-3717 (2022).

[87] Plutchik, R.: The Nature of Emotions: Clinical Implications, in Emotions and Psychopathology, Clynes, M. and Panksepp, J. (eds), Springer, pp.1-20 (1988).

[88] Fugate, J. M. B. and Franco, C. L.: What Color Is Your Anger? Assessing Color-Emotion Pairings in English Speakers, Frontiers in Psychology, Vol.10, Article No.206 (2019).

[89] Gadde, V. and Beykpour, K.: An update on our work around the 2020 US Elections, X Blog (online), available from <https://blog.twitter.com/en_us/topics/company/2020/2020-election-update> (accessed 2024-12-23).

[90] X blog: Helpful Birdwatch notes are now visible to everyone on Twitter in the US, X blog (online), available from <https://blog.twitter.com/en_us/topics/product/2022/helpful-birdwatch-notes-now-visible-everyone-twitter-us> (accessed 2024-12-23).

[91] Jigsaw: Using machine learning to reduce toxicity online, Jigsaw (online), available from <https://perspectiveapi.com> (accessed 2024-12-23).